\def\text#1{{\rm{#1}}}

\def\X{{\cal{X}}}

\def\Tr{\text{Tr\:}}
\def\>{\rangle}
\def\<{\langle}

\def\2{{\frac{1}{2}}}

\def\Det{{\rm{Det}}}
\def\1{{{-1}}}

\def\P{{\partial}}
\def\bP{{\bar\partial}}
% differential forms

% partial right

% partial left

% partial leftright

% partial slash

% partial slash right

% partial slash left

\def\a{{\alpha}}
\def\b{{\beta}}
\def\g{{\gamma}}
\def\d{{\delta}}

\def\t{{\theta}}
\def\w{{\omega}}

\def\r{{\rho}}
\def\s{{\sigma}}
\def\m{{\mu}}
\def\n{{\nu}}
\def\k{{\kappa}}
\def\l{{\lambda}}

\def\newpage{\vfill\eject}

\input harvmac
\input epsf

\def\frac#1#2{{#1\over#2}}

\font\ensX=msbm10
\font\ensVII=msbm7
\font\ensV=msbm5
\newfam\math
\textfont\math=\ensX \scriptfont\math=\ensVII \scriptscriptfont\math=\ensV
\def\ensemble{\fam\math\ensX}

% add a caption. The first argument must be a length
\def\caption#1#2{\leftskip=#1\rightskip=#1\noindent\ninepoint\it #2 % the following blank line is important

\leftskip=0cm\rightskip=0cm\tenpoint}

%%%%%%%%%%%%%%%%%%%%%%%%%%%%%%%%%%%%%%%%%%%%%%%%%%%%%%%%
%
%          REFERENCES
%
%%%%%%%%%%%%%%%%%%%%%%%%%%%%%%%%%%%%%%%%%%%%%%%%%%%%%%%%

\lref\McGreevyTV{
%\bibitem{McGreevy:2003ep}
  J.~McGreevy, J.~Teschner and H.~L.~Verlinde,
  ``Classical and quantum D-branes in 2D string theory,''
  JHEP {\bf 0401} (2004) 039
  [arXiv:hep-th/0305194].
  %%CITATION = HEP-TH 0305194;%%
}

\lref\McGreevyV{
%\bibitem{McGreevy:2003kb}
  J.~McGreevy and H.~L.~Verlinde,
  ``Strings from tachyons: The c = 1 matrix reloaded,''
  JHEP {\bf 0312}, 054 (2003)
  [arXiv:hep-th/0304224].
  %%CITATION = HEP-TH 0304224;%%
  }

\lref\Klebanov{
%\bibitem{Klebanov:2003km}
  I.~R.~Klebanov, J.~M.~Maldacena and N.~Seiberg,
  ``D-brane decay in two-dimensional string theory,''
  JHEP {\bf 0307} (2003) 045
  [arXiv:hep-th/0305159].
  %%CITATION = HEP-TH 0305159;%%
  }
  
%\ReinschYK  
\lref\ReinschYK{
  M.~W.~Reinsch,
  ``A Simple expression for the terms in the Baker-Campbell-Hausdorff series,''
  J.\ Math.\ Phys.\  {\bf 41}, 2434 (2000)
  [arXiv:math-ph/9905012].
  %%CITATION = MATH-PH 9905012;%%
}

%\MarcoDiplom
\lref\MarcoDiplom{
  M.~Baumgartl,
  ``On Boundary States in Superstring Theory,''
  diploma thesis.
}

%\BaumgartlIY
\lref\BaumgartlIY{
  M.~Baumgartl, I.~Sachs and S.~L.~Shatashvili,
  ``Factorization Conjecture And The Open / Closed String Correspondence,''
  JHEP {\bf 0505}, 040 (2005)
  [arXiv:hep-th/0412266].
  %%CITATION = HEP-TH 0412266;%%
}

%\Fradkin
\lref\Fradkin{
E. ~S.~Fradkin and A.~A.~Tseytlin, "Quantum String Theory Effective Action", 
Nucl.\ Phys.\ B {\bf 261}, 1 (1985).
}

%\TseytlinRR
\lref\TseytlinRR{
  A.~A.~Tseytlin,
  ``Sigma model approach to string theory,''
  Int.\ J.\ Mod.\ Phys.\ A {\bf 4}, 1257 (1989).
  %%CITATION = IMPAE,A4,1257;%%
}

%\WittenQY
\lref\WittenQY{
  E.~Witten,
  ``On background independent open string field theory,''
  Phys.\ Rev.\ D {\bf 46}, 5467 (1992)
  [arXiv:hep-th/9208027].
  %%CITATION = HEP-TH 9208027;%%
}

%\WittenCR
\lref\WittenCR{
  E.~Witten,
  ``Some computations in background independent off-shell string theory,''
  Phys.\ Rev.\ D {\bf 47}, 3405 (1993)
  [arXiv:hep-th/9210065].
  %%CITATION = HEP-TH 9210065;%%
}

%\ShatashviliKK
\lref\ShatashviliKK{
  S.~L.~Shatashvili,
  ``Comment on the background independent open string theory,''
  Phys.\ Lett.\ B {\bf 311}, 83 (1993)
  [arXiv:hep-th/9303143].
  %%CITATION = HEP-TH 9303143;%%
}

%\ShatashviliPS
\lref\ShatashviliPS{
  S.~L.~Shatashvili,
  ``On the problems with background independence in string theory,''
  Alg.\ Anal.\  {\bf 6}, 215 (1994)
  [arXiv:hep-th/9311177].
  %%CITATION = HEP-TH 9311177;%%
}

%\GreenSP
\lref\GreenSP{
  M.~B.~Green, J.~H.~Schwarz and E.~Witten,
  ``Superstring Theory. Vol. 1: Introduction,''
	%\href{http://www.slac.stanford.edu/spires/find/hep/www?irn=1755021}{SPIRES entry}
}
%\GreenMN
\lref\GreenMN{
  M.~B.~Green, J.~H.~Schwarz and E.~Witten,
  ``Superstring Theory. Vol. 2: Loop Amplitudes, Anomalies And Phenomenology,''
 %\href{http://www.slac.stanford.edu/spires/find/hep/www?irn=1782614}{SPIRES entry}
}

%\ZwiebachFE
\lref\ZwiebachFE{
  B.~Zwiebach,
  ``Oriented open-closed string theory revisited,''
  Annals Phys.\  {\bf 267}, 193 (1998)
  [arXiv:hep-th/9705241].
  %%CITATION = HEP-TH 9705241;%%
}

%\StromingerZD
\lref\StromingerZD{
  A.~Strominger,
  ``Closed strings in open string field theory,''
  Phys.\ Rev.\ Lett.\  {\bf 58}, 629 (1987).
  %%CITATION = PRLTA,58,629;%%
}

%\ShatashviliXXX
\lref\ShatashviliXXX{
  S.~Shatashvili,
  ``Closed strings as solitons in background independent open string field theory,''
  unpublished, talk at IHES, Paris, July 1997.
}

%\LiZA
\lref\LiZA{
  K.~Li and E.~Witten,
  ``Role of short distance behavior in off-shell open string field theory,''
  Phys.\ Rev.\ D {\bf 48}, 853 (1993)
  [arXiv:hep-th/9303067].
  %%CITATION = HEP-TH 9303067;%%
}

%\PressleyQK
\lref\PressleyQK{
A.~Pressley and G.~Segal,
``Loop Groups,''
Oxford, UK: Clarendon (1988) 318p. (Oxford Mathematical Monographies)
%\href{http://www.slac.stanford.edu/spires/find/hep/www?irn=1979159}{SPIRES entry}
}

%\MickelssonHP
%\lref\michII{
\lref\MickelssonHP{
J.~Mickelsson,
``Current Algebras And Groups,''
New York, USA: Plenum (1989) 313p.
%\href{http://www.slac.stanford.edu/spires/find/hep/www?irn=2184869}{SPIRES entry}
}

%\MickelssonSB
\lref\MickelssonSB{
%\lref\michI{
J.~Mickelsson,
``On The 2 Cocycle Of A Kac-Moody Group,''
Phys.\ Rev.\ Lett.\  {\bf 55}, 2099 (1985).
%%CITATION = PRLTA,55,2099;%%
}

%\SenNF
\lref\SenNF{
  A.~Sen,
  ``Tachyon dynamics in open string theory,''
  Int.\ J.\ Mod.\ Phys.\ A {\bf 20}, 5513 (2005)
  [arXiv:hep-th/0410103].
  %%CITATION = HEP-TH 0410103;%%
}
%\SenZM
\lref\SenZM{
  A.~Sen,
   ``Rolling tachyon boundary state, conserved charges and two dimensional string theory,''
  JHEP {\bf 0405}, 076 (2004)
  [arXiv:hep-th/0402157].
  %%CITATION = HEP-TH 0402157;%%
}

%\GaiottoYB
\lref\GaiottoYB{
  D.~Gaiotto and L.~Rastelli,
   ``A paradigm of open/closed duality: Liouville D-branes and the Kontsevich model,''
  JHEP {\bf 0507}, 053 (2005)
  [arXiv:hep-th/0312196].
  %%CITATION = HEP-TH 0312196;%%
}
%\ShatashviliUX
\lref\ShatashviliUX{
  S.~L.~Shatashvili,
   ``On field theory of open strings, tachyon condensation and closed
  strings,''
  arXiv:hep-th/0105076.
  %%CITATION = HEP-TH 0105076;%%
}
%\GaiottoRM
\lref\GaiottoRM{
  D.~Gaiotto, N.~Itzhaki and L.~Rastelli,
  ``Closed strings as imaginary D-branes,''
  Nucl.\ Phys.\ B {\bf 688}, 70 (2004)
  [arXiv:hep-th/0304192].
  %%CITATION = HEP-TH 0304192;%%
}
%\SenXS
\lref\SenXS{
  A.~Sen,
  ``Open-closed duality at tree level,''
  Phys.\ Rev.\ Lett.\  {\bf 91}, 181601 (2003)
  [arXiv:hep-th/0306137].
  %%CITATION = HEP-TH 0306137;%%
}
%\SenZM
\lref\SenZM{
  A.~Sen,
   ``Rolling tachyon boundary state, conserved charges and two dimensional string theory,''
  JHEP {\bf 0405}, 076 (2004)
  [arXiv:hep-th/0402157].
  %%CITATION = HEP-TH 0402157;%%
}
%\SenYV
\lref\SenYV{
  A.~Sen,
   ``Symmetries, conserved charges and (black) holes in two dimensional string theory,''
  JHEP {\bf 0412}, 053 (2004)
  [arXiv:hep-th/0408064].
  %%CITATION = HEP-TH 0408064;%%
}

%\AlekseevMC
\lref\AlekseevMC{
  A.~Y.~Alekseev and V.~Schomerus,
  ``D-branes in the WZW model,''
  Phys.\ Rev.\ D {\bf 60}, 061901 (1999)
  [arXiv:hep-th/9812193].
  %%CITATION = HEP-TH 9812193;%%
}

%\KlimcikHP
\lref\KlimcikHP{
  C.~Klimcik and P.~Severa,
  ``Open strings and D-branes in WZNW models,''
  Nucl.\ Phys.\ B {\bf 488}, 653 (1997)
  [arXiv:hep-th/9609112].
  %%CITATION = HEP-TH 9609112;%%
}

%\GawedzkiBQ
\lref\GawedzkiBQ{
  K.~Gawedzki,
  ``Conformal field theory: A case study,''
  arXiv:hep-th/9904145.
  %%CITATION = HEP-TH 9904145;%%
}

%\tHooftXB
\lref\tHooftXB{
  G.~'t Hooft, C.~Itzykson, A.~Jaffe, H.~Lehmann, P.~K.~Mitter, I.~M.~Singer and R.~Stora,
   ``Recent Developments In Gauge Theories. Proceedings, Nato Advanced Study Institute, Cargese, France, August 26 - September 8, 1979,''
%\href{http://www.slac.stanford.edu/spires/find/hep/www?irn=949701}{SPIRES entry}
}

%\GrossEG
\lref\GrossEG{
  D.~J.~Gross, A.~Neveu, J.~Scherk and J.~H.~Schwarz,
  ``Renormalization and unitary in the dual-resonance model,''
  Phys.\ Rev.\ D {\bf 2}, 697 (1970).
  %%CITATION = PHRVA,D2,697;%%
}
%\LovelaceFA
\lref\LovelaceFA{
  C.~Lovelace,
  ``Pomeron Form-Factors And Dual Regge Cuts,''
  Phys.\ Lett.\ B {\bf 34}, 500 (1971).
  %%CITATION = PHLTA,B34,500;%%
}
%\StromingerZD
\lref\StromingerZD{
  A.~Strominger,
  ``Closed strings in open string field theory,''
  Phys.\ Rev.\ Lett.\  {\bf 58}, 629 (1987).
  %%CITATION = PRLTA,58,629;%%
}

%\BonoraTM
\lref\BonoraTM{
  L.~Bonora, N.~Bouatta and C.~Maccaferri,
  ``Towards open-closed string duality: Closed strings as open string fields,''
  arXiv:hep-th/0609182.
  %%CITATION = HEP-TH 0609182;%%
}
%\KatsumataCC
\lref\KatsumataCC{
  F.~Katsumata, T.~Takahashi and S.~Zeze,
   ``Marginal deformations and closed string couplings in open string field
  theory,''
  JHEP {\bf 0411}, 050 (2004)
  [arXiv:hep-th/0409249].
  %%CITATION = HEP-TH 0409249;%%
}

%\GarousiPI
\lref\GarousiPI{
  M.~R.~Garousi and G.~R.~Maktabdaran,
  ``Closed string S-matrix elements in open string field theory,''
  JHEP {\bf 0503}, 048 (2005)
  [arXiv:hep-th/0408173].
  %%CITATION = HEP-TH 0408173;%%
}
%\SenIV
\lref\SenIV{
  A.~Sen,
  ``Open-closed duality: Lessons from matrix model,''
  Mod.\ Phys.\ Lett.\ A {\bf 19}, 841 (2004)
  [arXiv:hep-th/0308068].
  %%CITATION = HEP-TH 0308068;%%
}
%\DiVecchiaAE
\lref\DiVecchiaAE{
  P.~Di Vecchia, A.~Liccardo, R.~Marotta and F.~Pezzella,
  ``Gauge / gravity correspondence from open / closed string duality,''
  JHEP {\bf 0306}, 007 (2003)
  [arXiv:hep-th/0305061].
  %%CITATION = HEP-TH 0305061;%%
}
%\MayrXK
\lref\MayrXK{
  P.~Mayr,
  ``N = 1 mirror symmetry and open/closed string duality,''
  Adv.\ Theor.\ Math.\ Phys.\  {\bf 5}, 213 (2002)
  [arXiv:hep-th/0108229].
  %%CITATION = HEP-TH 0108229;%%
}
%\KapustinDF
\lref\KapustinDF{
  A.~Kapustin and L.~Rozansky,
  ``On the relation between open and closed topological strings,''
  Commun.\ Math.\ Phys.\  {\bf 252}, 393 (2004)
  [arXiv:hep-th/0405232].
  %%CITATION = HEP-TH 0405232;%%
}
%%%%%%%%%%%%%%%%%%%%%%%%%%%%%%%%%%%%%%%%%%%%%%%%%%%%%%%%
%
%            BODY
%
%%%%%%%%%%%%%%%%%%%%%%%%%%%%%%%%%%%%%%%%%%%%%%%%%%%%%%%%

%%%%%%%%%%%%%%%%%%%%%%%%%%%%%%%%%%%%%%%%%%%%%%%%%%%%%%%%%%%%%%%%%%%%
\Title{\vbox{\baselineskip12pt
		\hbox{hep-th/0611112}
		\hbox{LMU-ASC 80/06}
}}{\vbox{
\centerline{Open-closed string correspondence:}
\vskip2pt
\centerline{D-brane decay in curved space}
 }}

%\medskip
\centerline{\bf M. Baumgartl\footnote{$^\star$}{E-mail: {\tt baumgartl@itp.phys.ethz.ch}} 
	and I. Sachs\footnote{$^\dagger$}{E-mail: {\tt ivo@theorie.physik.uni-muenchen.de}}}
\bigskip\bigskip
\centerline{\it $^{\star}$ 
Institute for Theoretical Physics, ETH Z\"urich}
\centerline{\it 8093 Zurich, Switzerland}
\medskip
\centerline{\it $^{\dagger}$ 
Arnold Sommerfeld Center for Theoretical Physics, LMU M\"unchen}
\centerline{\it
Theresienstr. 37, 80333 Munich, Germany}
\centerline{and}
\centerline{\it Department of Physics and Astronomy, UCLA}
\centerline{\it Los Angeles, CA 90095-1547, USA}
\vskip 1cm

This paper analyzes the effect of curved closed string backgrounds on the stability of D-branes within boundary string field theory. We identify the non-local open string background that implements shifts in the closed string background and analyze the tachyonic sector off-shell. The renormalization 
group flow reveals some characteristic properties, which are expected for 
a curved background, like the absence of a stable space-filling brane. 
In 3-dimensions we describe tachyon condensation processes to lower-dimensional branes, 
including a curved 2-dimensional brane. We argue that this 2-brane is perturbatively 
stable. This is in agreement with the known maximally symmetric WZW-branes and 
provides further support to the bulk-boundary factorization approach to 
open-closed string correspondence.

\Date{}

\newpage

\newsec{Introduction: partition function approach to string field theory}

Critical string theory is usually described by a conformal field theory, representing the string worldsheet embedded in target space. The requirement that conformality conditions hold is in general enough to provide a definition of critical string theory in different target spaces. In fact, any target space, for which a conformal field theory can be written down should be a solution of general string field theory equations of motion.

It is an important issue to understand, how these different critical theories are related, or, rephrased similarly vaguely, how the space of string theories can be parametrized off-shell. An approach which has been followed since the early days of string theory is to extend the space of two-dimensional conformal theories to the space of two-dimensional field theories with arbitrary couplings. This approach proved to be rather successful in the open string sector \refs{\Fradkin\TseytlinRR\WittenQY\WittenCR\ShatashviliKK{--}\ShatashviliPS}. 
The prime situation studied has been bosonic string theory on a flat closed string background with arbitrary tachyon and photon boundary perturbations. In this case the agreement between a generalized $\sigma$-model approach and on-shell S-matrix scattering has been proven.

In \refs{\WittenQY\WittenCR\ShatashviliKK{--}\ShatashviliPS} a class of generalized $\sigma$-model has been considered which has been obtained from conformal field theories by the inclusion of arbitrary (possibly non-renormalizable) interaction terms on the boundary of the worldsheet. Under some assumptions the partition function approach was then extended to provide a background independent, off-shell, open string field theory action formulated in terms of space-time  fields. Roughly, there are two types of input data necessary for the construction of the action: first, the bulk conformal field theory must be specified, which corresponds to the choice of a closed string background. As there is no complete classification of conformal field theories, the space of `closed string backgrounds' is not well-defined\foot{See \ZwiebachFE\ for attempts in closed string field theory.}. Equally poorly understood is the space of boundary interaction terms, which is used to deform the boundary conformal field theory. This space is intimately related to the configuration space of open string field theory.

On the other hand, these two spaces are certainly not independent of each other, because some examples of dualities between open and closed strings are known 
\refs{\GrossEG\LovelaceFA\StromingerZD\BonoraTM\KatsumataCC\GarousiPI\SenIV\DiVecchiaAE\ShatashviliUX\MayrXK{--}\KapustinDF}. These correspondences are usually observed in some special situations, but methods for a systematic investigation are still lacking. Depending on one's point of view one could try to start with open string field theory, add closed strings by hand and divide out multiplicities in the scattering amplitudes \ZwiebachFE. The alternative approach, which we will follow, is to identify the collective open string excitations which correspond to deformations of the closed string background \refs{\StromingerZD,\ShatashviliXXX}. The problem with this approach is that the `space of open string interactions' also has to be defined first\foot{The situation is under reasonable control in 2d string theory where a non-local map between closed string degrees of freedom and open string excitations can be constructed \refs{\McGreevyTV\McGreevyV{--}\Klebanov}. There, the closed string excitations are given by the excitations of the open string `Fermi sea'.}. The usual truncation to massless and tachyonic modes is certainly too restrictive in this case. On the other hand, if infinitely many boundary couplings are present, one could expect that some of them can be summed up to give non-local couplings \LiZA. This possibility will be explored here in the context of the recent bulk-boundary factorization conjecture. In \BaumgartlIY\ it was argued that the partition functions for generalized boundary conformal field theories factorize into a bulk part and a boundary partition function provided the bulk is kept on-shell. 

For the definition of an open string field theory only the boundary partition function is of importance. Factorization then implies that two open string field theory actions constructed initially over different closed string backgrounds, can be mapped to each other explicitly provided one can find a field re-definition between the two target spaces, at least locally. This has been done in \BaumgartlIY, where it was found that there are universality classes of open string couplings labeled by non-local, collective open string boundary perturbations. It was conjectured that these universality classes correspond to distinct closed string backgrounds.

In this note we give further support to the idea advocated in \BaumgartlIY, that in open string field theory, closed string backgrounds are equivalent to certain collective open string excitations by investigation of a concrete example. We compare certain off-shell properties of open string field theory in flat space to those of strings propagating on a $SU(2)^{\ensemble{C}}$ group manifold. We work perturbatively in the large radius limit and we focus on the existence of D-branes which serves as a check for the conjectured correspondence. We observe some generic effects which are expected based on what is known about conformal D-branes in $SU(2)$ WZW-models. Concretely, we will identify certain D-branes in the non-locally deformed boundary CFT and find agreement with the maximally symmetric D-branes in $SU(2)$. The non-local open string background (in the large radius limit) is interpreted as a deformation of flat space towards a curved space, resulting in renormalization group flow that destabilizes the space-filling brane through decay into a lower dimensional brane.

The plan is as follows: we introduce the model in the next section and construct the boundary action for open strings propagating on a group manifold. We then show how to apply the general construction to the $SU(2)$ target space. In section 3 we carry out the renormalization and compute $\beta$-functions on a space filling brane. Section 4 discusses the renormalization group flow in terms of tachyon condensation on the space filling brane. In the 5th section we give further support for the condensation scenario proposed in section 4 by verifying that the curved 2-branes is indeed a perturbative fixed point of the RG-flow. Section 6 contains our conclusions.

\newsec{Boundary action from bulk/boundary factorization}
In boundary string field theory one seeks to define an action \refs{\WittenQY\WittenCR\ShatashviliKK{--}\ShatashviliPS} on 
the space of $\sigma$-models with boundary. In the general case the 
definition of such an action is an open problem. However, when ghost and 
matter fields decouple the space-time action can be defined as \refs{\ShatashviliKK,\ShatashviliPS}
\eqn\BSFTA{\eqalign{
	S[t^I]&=(1-\beta^I\partial_I)Z_{bdry}[t^I]
}}
where $Z_{bdry}[t_I]$ is the boundary partition function with boundary interactions parametrized by $t^I$. For instance, for free bosons taking values on a torus with radius $R$, this function is given by the integral over all boundary fields $X_b(z,\bar z)= f(z)+\bar f(\bar z)+ f_0$
\eqn\BSFTB{\eqalign{
	Z_{bdry}[t^I] &= \int D[X_b] e^{-\frac{1}{\alpha`}I_{bdry}[t^I,X_b]}
}}
with boundary action
\eqn\BSFTC{\eqalign{
	I_{bdry}[t^I,X_b] &= R^2\oint f(\theta)\partial_\theta \bar f(\theta) +V(t^I,X_b(\theta))
}}
where $V(t^I,X)$ stands for all possible boundary interactions given by its expansion in terms of $X_b$, i.e. 
 \eqn\BSFTD{\eqalign{
	V[t^I,X_b] &= T(X_b) +A_\mu(X_b)\partial_\theta X_b^\mu+ \dots,
}}
where $\left\{t^I\right\}=\left\{T, A, \dots\right\}$.
The first term in $I_{bdry}$ is a non-local expression in $X_b$ that arises when pulling the 2d-$\sigma$ model action back to the boundary of the worldsheet. In this section we will show how the non-locality in $I_{bdry}$ is modified when the flat closed string background is replaced by a curved one\foot{The closed string background is always assumed to be conformal.}. 
 
The factorization property of the partition function on curved backgrounds is by no means obvious but it was shown to hold in \BaumgartlIY\ when the target space is a group manifold. In the present paper, for concreteness, we will take as a starting point a boundary theory with $S^3$ target, that is, the WZW on $SU(2)$. It was shown in \BaumgartlIY\ that for this model the partition function factorizes into a bulk contribution $Z_0$ and a boundary contribution $Z_{bdry}[t^I]$. Let us briefly recapitulate the arguments leading to the expression for the boundary action, starting with the general case of an unspecified Lie group.

\subsec{Construction of the boundary action for Lie groups}

Consider a WZW model with fields $g$ taking values in group 
group $G$. We write the action as 
\eqn\EQNWZW{\eqalign{
	I[g] &= L[g] + \Gamma^\b[g],
}}
where $L[g]$ is the kinetic term and $\Gamma[g]$ is the topological term, 
which can be written as an integral of a closed $3$-form in target-space. 
For this, the group valued fields $g:\Sigma\to G$, 
which parametrize the embedding of the 2-dimensional world sheet 
in the group, must be extended to $\tilde g:B\to G$, where $\Sigma=\P B$. 
For $\Sigma$ closed this procedure produces a well-defined partition function. If $H^3(G)=0$ then, if the world sheet has a boundary, $\Gamma$ can still be defined modulo an ambiguity which is parametrized by a boundary integral\foot{If $H^3=Z$ the definition of the boundary WZW model is problematic 
\GawedzkiBQ, however it was argued in \BaumgartlIY\ that the final result for the boundary partition function will still hold. In the case at hand we will soon see that the boundary model is really defined on some 3-dimensional, non-compact hypersurface in the complexified group $SL(2)^{\ensemble C}$.}, $\oint \b$, of a $1$-form $\b\in\Omega^1(G)$. Upon expansion of $\beta$ in terms of derivatives of the fields one then recovers the open sting degrees of freedom inserted at the boundary of the world sheet. Thus the ambiguity in the choice of $\beta$ reflects the choice of an open string background. 
For arbitrary $\b$ the boundary theory will not be scale invariant, but we will argue below that for a given closed string background there exists a $\beta$ such that the theory is conformal.

The bulk-boundary factorization procedure relies on splitting the group-valued field $g$ in two parts, $g_0$ and $k$, so that the action is a functional of the product $g=g_0k$.  The `$D0$ part' $g_0$ satisfies Dirichlet boundary conditions in all directions, i.e. $g_0|_{\partial\Sigma}\equiv 1$. The classical part $k(z,\bar z)=h(z)\bar h(\bar z)$ obeys the equations of motion by construction and is thus uniquely determined by the boundary conditions on $g$. 

Note that not every field $g$ can be split in such a manner. For instance Neumann boundary conditions cannot be obtained as a deformation of Dirichlet boundary conditions. However, one obtains Neumann boundary conditions upon integration over $k$ \MarcoDiplom.

With this Ansatz for $g$ the $\sigma$-model action then decomposes into
\eqn\EQNbaI{\eqalign{
	I(g=g_0k) &= I(g_0)+L(k)+\Gamma^\b(k) + W(g_0,k)
}}
as a consequence of the Polyakov-Wiegmann formula\foot{The Polyakov-Wiegmann formula is valid for worldsheets with boundary. Its appropriate generalization for boundary $\s$-model was discussed in \BaumgartlIY.}. Moreover is was shown, that the mixed term $W(g_0,k)$ does not contribute to boundary correlation functions at any order in perturbation theory by explicit computation. This finally allows one, from a BSFT point of view, to consider the boundary partition function for the action $I(k)=L(k)+\Gamma^\b(k)$ which is manifestly independent of the bulk fields.

Here we are interested in concrete calculations in order to clarify the effect of deformations of the closed string backgrounds on the RG flows and its fixed points which characterize conformal boundary conditions. Hence an explicit expression for $I(k)$ is required. After choosing a parametrization of the group, the boundary action \EQNbaI\ can be written down easily. The only term which needs some attention is the topological term $\Gamma^\beta=\int w_3$. Locally, $w_3=dw_2$. We then observe that
\eqn\EQNCOC{\eqalign{
	\g(h,\bar h) \equiv w_2(h\bar h)-w_2(h)-w_2(\bar h)+\tr h^{-1}dh d\bar h\bar h^{-1}
}} 
is a well-defined closed two-form in $\Omega^2(\Sigma)$, which furthermore satisfies the two-cocycle condition. Since fields $h$ and $\bar h$ are holomorphic and anti-holomorphic respectively we have in fact
\eqn\X{\eqalign{
	\g(h,\bar h) &= w_2(h\bar h)+\tr h^{-1}\P h \bP \bar h\bar h^{-1} \cr
				&= w_2(k) + \tr k^{-1}\P kk^{-1}\bP k.
}}
$\g$ is an expression for the modification of the Polyakov-Wiegmann formula applied to $\Gamma$ on a worldsheet with boundary. Its integral
\eqn\X{\eqalign{
	\a(h,\bar h) \equiv \int_{\Sigma} \g(h, \bar h).
}}
over a two-manifold depends only on the boundary data. Thus $\alpha(h,\bar h)$ defines a group 2-cocycle \refs{\PressleyQK\MickelssonHP{--}\MickelssonSB}. 
The topological term $\Gamma^\beta$ can thus be written as 
\eqn\X{\eqalign{
	\Gamma^\beta(k) = -L(k) + \a(h,\bar h),
}}
modulo the ambiguity parametrized by the 1-form $\beta(k)$. Adding the kinetic term we find that the boundary action is just given by $\a(h,\bar h)$ modulo the overall ambiguity $\oint \b$.

It is tempting to speculate that in BSFT the space of $d$-dimensional closed string backgrounds can be classified in terms of the deformations of $U(1)^d$ group cocycles but we shall not elaborate further on this idea in the present paper. 

In order to find a representative for $\a$ we will make a natural choice and show that it is consistent with several properties imposed on the boundary action: It must reproduce the flat space model in the large-radius limit; it must lead to the same algebraic boundary conditions as the full WZW model; and it must satisfy the cocycle condition. A choice for the boundary action which satisfies this is
\eqn\EQNgamma{\eqalign{
	\g(h,\bar h) &=L(k) = h^{-1}\P h \bP\bar h\bar h^{-1}. 
}} 
Next we show, that the claimed properties are satisfied.

1. In order to conduct the large-radius limit, we choose the standard parametrization for Lie-groups by
\eqn\X{\eqalign{
	h(z) &= \exp \frac{i f^\mu(z)T_\mu}{\sqrt\k} \qquad \bar h(\bar z) = \exp \frac{i\bar f^\m (\bar z)T_\m}{\sqrt\k}
}} 
with $\bar f_n=f_n^*$. $T_\mu$ are the generators of the group. At $\k=\infty$ it reduces to $\oint f\bP\bar f$, so the group coordinates $h$ and $\bar h$ can be related to the flat space coordinates $f$ and $\bar f$ by an expansion around the infinite radius point.

It is instructive to consider how `flat coordinates' and the `curved split coordinates' $h$, $\bar h$ are connected in this construction. First, $\ln k$ can be expressed in terms of $\ln h$ and $\ln \bar h$ by Baker-Campbell-Hausdorff expansion (see \ReinschYK\ for a nice treatment). Thus $\ln k\sim f+\bar f + $ corrections. The corrections contribute with $\kappa^{-\frac{1}{2}}$, so they are come from a slight deviation from flat geometry. In the limit $\k\to\infty$ we can make contact with the standard boson on $T^3$ described in coordinates $X$ by setting $X=f+\bar f$. Then we see that the next order in $\kappa^{-\frac{1}{2}}$-expansion gives a correction to the flat coordinates,
\eqn\EQNCORR{\eqalign{
	X^\m \approx X^\m_{(0)} + \frac{1}{\sqrt\k} X^\m_{(1)}&= f^\m + \bar f^\m - \frac{1}{\sqrt\k}\epsilon^{\m\n\l}f_\n\bar f_\l.
}}

The `flat coordinates' $X_{(0)}$ obey a reality constraint enforcing $f^*=\bar f$, which might be somewhat unnatural from the WZW point of view since then $k$ is not unitary or, equivalently, $X^\mu$ is not real. Nevertheless utilizing the reality  constraint for $f$ one finds that the action is real (essentially because it reduces to a boundary integral). The reality constraint in the flat coordinates then implies that the boundary theory is defined not on $SU(2)$, but on its complexification $SL(2)^{\ensemble C}$. This is not done by doubling the degrees of freedom, because the complex corrections are determined in terms of $f$ and $\bar f$.

2. Boundary conditions are obtained from the Lagrangian in form of a surface term in the variation with respect to the fields. In the full WZW model we will get two terms, $\d L$ and $\d\Gamma$. But $\d\Gamma$ produces no surface term, rather it contributes only to the equations of motion. Therefore boundary conditions must be completely determined by $\d L$. Evaluated on classical solutions, it is exactly given by \EQNgamma, which is a pure surface term. Therefore the boundary Lagrangian \EQNgamma\ must reproduce the same boundary conditions.

3. Finally we show that our choice of $\g$ indeed is a cocycle. For this we need to show that
\eqn\X{\eqalign{
	\gamma(a,b)+\gamma(ab,c) &= \gamma(b,c) + \gamma(a,bc).
}}
Here $a,b,c \in G(z,\bar z)$ are arbitrary functions over $G$. Direct computation reveals, that the cocycle condition is satisfied.
Additionally $\g$ is a closed 2-form on the worldsheet.

From this we conclude that $\g=h^{-1}\P h\bP\bar h\bar h^{-1}$ is a valid representative for the cocycle and can serve as a boundary action.

\subsec{The $SU(2)$ boundary action}
\subseclab\SECBOUNDARYACTION

We now specify the closed string background to be the group manifold of $SU(2)$ (or its complexification) and set the one-form $\b$ to zero temporarily.
For convenience we substitute $\l\equiv\k^{-\frac{1}{2}}$ and obtain the $SU(2)$-boundary action\foot{The trace is normalized in a way so that the quadratic part of the action is given by the standard term $\sum_{m>0}mf_m\bar f_m$ in flat space.}
\eqn\X{\eqalign{
	S=\frac{1}{(i\l)^2}\Tr \int \P kk^{-1}\bP kk^{-1}.
}}
Expanding the action up to fourth order in $f$ and $\bar f$ one finds after some tedious algebra
\eqn\EQNAC{\eqalign{
	S = s\sum_{m=1}m&f_m^\a\bar f_{m\a}
		+ \a (V_\a - V_{\bar\a}) + \b V_\b + \g (V_\g + V_{\bar\g}),
}}
with
\eqn\EQNACII{\eqalign{
	V_\a &= \sum_{c,b,a=1}(b-a)\d_{c,a+b} \epsilon_{\m\n\l}   f^\m_a f^\n_b  \bar f^\l_c\cr
	V_\b &= - \sum_{a,b,c,d=1}\frac{(c-b)(a-d)}{a+d} \d_{c+b,a+d} f^\m_c  f^\n_b \bar f_{a\m}  \bar f_{d\n} \cr
	V_\g &= - \frac{2}{3}\sum_{a,b,c,d=1}(a-d-b)\d_{c,b+a+d}
			   f^\m_c \bar f_{a\m} \bar f^\n_b \bar f_{d\n},
}}
where $V_{\bar\a}$ and $V_{\bar\g}$ is obtained by exchanging $f$ with $\bar f$. The original action (the starting point of the renormalization group flow) is found for $s=1, \a=\frac{\l}{2}, \b=\g=\frac{\l^2}{2}$.
Although the action can be computed exactly we truncate its expansion at ${\cal O}(\l^3)$ (we will see that this gives $\beta$-functions which are exact up to ${\cal O}(\l^5)$).

There is another contribution to the action coming from a Jacobian due to the change of variables. Starting from the standard Haar measure on the group $[dk]=k^{-1}dk$ we obtain the following Jacobian:
\eqn\X{\eqalign{
	J &= \frac{\left(\left [\d kk^{-1}\right ]^+, \left [\d kk^{-1}\right ]^-\right)}{\left(\d f, \d\bar f\,\right)}
	= \pmatrix{ \frac{\left [\d hh^{-1}\right ]^+}{\vphantom{\hat f}\d f}\quad \, 
				\frac{\left [h\d \bar h\bar h^{-1}h^{-1}\right ]^+}{\vphantom{\hat f} \d \bar f}\cr
		 \qquad 0\qquad\, \frac{\left [h\d \bar h\bar h^{-1}h^{-1}\right ]^-}{\vphantom{\hat f}\d \bar f} }
		= \pmatrix{\;J_{11}\;\,J_{12}\;\cr \;0\,\; \;\;J_{22}},
}}
where $\pm$ indicates restriction to the holomorphic/anti-holomorphic part.
Thus $\Det J=\Det J_{11} \cdot \Det J_{22}$. The rather lengthy calculation can be found in the appendix. Here we note simply the result. The measure contributes
\eqn\X{\eqalign{
	I_\text{measure} &= 4\l^2\sum_{n=1}f_{n\mu}\bar f^\mu_n + {\cal O}(\l^4)
}}
to the action. 
This is  a mass term for the boundary field $X_b$ which, due to its classical dimension, flows to $\infty$ in the infrared thus forcing $X_b$ to zero, i.e.  Dirichlet boundary conditions in all directions. Thus we see that  the tachyonic decay of the space-filling brane in the $SU(2)$ WZW-model is already encoded in the measure\foot{In fact, the measure for the boundary field $k$ is not uniquely determined by the bulk theory. Here we have taken the Haar measure 
as the starting point. Alternatively, one could consider the flat symplectic measure for $f$ and $\bar f$. In this case the tachyon arises as a one-loop counter term (see below).}.

\newsec{Renormalization of the $SU(2)$ boundary action}

Once the concrete form of the action has been obtained, we can analyze the quantized theory. It is clear that the action as it stands is not scale-invariant due to the presence of the mass term. To account for the mass and the `cosmological constant' we introduce the
tachyon coupling $T(X) = a + uf\bar f$. Note that we do not expect the expansion up to fourth order to be lead to a renormalizable theory. The exact action should, however, be renormalizable since the bulk theory from which it has been obtained is renormalizable. In particular we expect the renormalized action to describe field configurations in the same group manifold. Thus the structure of the interaction terms (which respects the group symmetry) should be untouched. Therefore we will assume $\l=\k^{-\frac{1}{2}}$ to be scale dependent, but keep the relative couplings fixed. Accounting for wave-function renormalization we will allow for $s$ to be scale dependent. 

For the calculation of the $\b$-functions we evaluate $n$-point functions expanded in loops. These correlators are IR finite because the theory is considered on a one-dimensional compact space. For large momentum the amplitudes are typically divergent, thus it is convenient to introduce a momentum cutoff $\Lambda$. This regularization seems appropriate as we are dealing with discrete sums so that $\Lambda$ simply appears as upper bound. The divergent parts of diagrams can be found by investigating the behavior for large $\Lambda$. Higher loop diagrams are treated in the following way. All loops naturally appear with sums over positive momenta only. Therefore sums of the type $\sum_{a=1}^\Lambda\sum_{b=1}^\Lambda f(a,b)$ can be transformed into an expression of the form $\sum_{\mu=2}^\Lambda\sum_{\nu=1}^{\mu-1} f(\nu,\mu-\nu)$. With this method only one divergent sum appears, even for higher loops.

Once the divergent part is extracted the renormalization procedure can be performed. Here we decide to start from the normal ordered theory with respect to the free field vacuum\foot{Such a normal ordering prescription can in general not be held at higher loops. In the approximation used here, however, it does hold, because all nested singularities of higher-loop diagrams ($\ge 2)$ are already removed through the 1-loop counter-terms. It turns out that this is not due to cancellations between different diagrams, but all diagrams become finite separately. The inclusion of self-contractions would only modify some numeric coefficients in the 1-loop counter-terms, which does not influence the finiteness of the 2-loop diagrams. The 3-loop-diagrams on the other hand vanish identically.}
and add counter-terms, which cancel the divergent part of the amplitudes. The counter-terms for the two- and three-point-functions ($p$ is the external momentum) are given by the following expression, which must be subtracted from the classical action: 
\eqn\EQNCOUNTER{\eqalign{
	\Sigma^{(2)}(p,\Lambda)
		&= \Lambda\left\{32\frac{\a^2}{s^2}\right\}
			+ \ln\Lambda\left\{-96 p\frac{\a^2}{s^2}-64\frac{u}{s}\frac{\a^2}{s^2}\right\} \cr
	\Sigma^{(3)}(p,p',\Lambda) 
		&=
			-(p'-p)\frac{4\a}{3s^2}(4\g-3\b) \ln\Lambda.
	}
}

Here, contributions up to three-loop order must be taken into account (although the 2- and 3-loop-contributions turn out to vanish). Now the $\b$-functions follow from \EQNCOUNTER, the conaonical dimensions of $a$ and $u$ as well as the vacum energy for free fields :
\eqn\EQNBETA{\eqalign{
	\matrix{
	\b_s &=& -96\frac{\a^2}{s^2} &\qquad \b_a &=& -a-\frac{u}{s} \cr
	\b_u &=& -u-64\frac{u\a^2}{s^3} &\qquad \b_\a &=& -\frac{4}{3}\frac{\a}{s^2}(4\g-3\b).
	}
}}   
The non-local couplings do not contribute a counter-term for the cosmological constant. Therefore the $\b$-function for $a$ is not modified by $\l$ and takes its usual form.

All these terms are one-loop contributions and therefore scheme independent. The two- and three-loop contribution to these $\beta$-functions vanish.
After absorption of the coupling $s$ into the field normalization and setting $\a=\frac{\l}{2}, \g=\b=\frac{\l^2}{2}$, the $\b$-functions become
\eqn\X{\eqalign{
	\b_{a} &= -a-u \cr
	\b_{u} &= -u (1- 8{\l}^2) \cr
	\b_{\l} &= -\frac{47}{6}\l^3 
}}
From these equations we can draw the following conclusions:

1. The coupling $\l$ for the non-local interaction increases under the renormalization group flow. This should not be taken as an indication that the curvature of the bulk background increases since the bulk theory, which is decoupled, is always on-shell. This coupling should rather be interpreted as an `auxiliary' coupling which mimics the effect of the closed string background on the open string dynamics. 

2. Tachyon condensation inevitably takes place. As we have seen above the tachyon is non-zero from the beginning due to the contribution of the measure. Furthermore, even if it were set to zero by an appropriate choice of the measure, a tachyon would be generated due to the one-loop counter-term. 

3. The running of $u$ is modified in a curved background. The way how $\l$ enters in $\b_u$ indicates that the tachyon flow in this example has a richer structure than in flat space.

At this point one could wonder about the end-point of the condensation. At perturbative level and with a finite set of couplings taken into account it is not possible to make definite predictions. The obtained $\b$-functions suggest that condensation to lower-dimensional branes can take place, in the same way as in flat space tachyon condensation. An infinite $u$ forces $f^\m\bar f_\m$ to zero, so that the resulting model will describe a D0-brane. The existence of a D0-brane is expected because it also arises in the WZW model (and is therefore compatible with the symmetries of the space). However, it is also possible identify a condensation process towards a higher-dimensional brane as endpoint. We present evidence for this in the next section.

\newsec{Tachyon condensation on the 3-brane}

The $\b$-functions \EQNBETA\ exhibit a complicated RG pattern, from which information about possible endpoints of the flow can be deduced. The trivial conformal point is $a=0, u=0, \l=0$, which is just the free boson theory without tachyon. Another well-known fixed point is obtained through tachyon condensation at $a=u=\infty$ (the Zamolodchikov metric vanishes at this point). We want to argue that there is another fixed point, which corresponds geometrically to a 2-brane. This must be expected from the study of D-branes in the WZW model \refs{\AlekseevMC,\KlimcikHP}.

In order to arrive at this conclusion, it is helpful to consider the boundary action with a tachyon insertion given by
\eqn\X{\eqalign{
\oint \beta(X_b) &=	\oint \r(X_b^2-c^2)^2.
}} 
In the case of finite $c$, condensation of $\r$ will lead to a localization on a spherical submanifold. Expansion up to third order in the fields yields an interaction term
\eqn\X{\eqalign{
	- 4\r c^2 f\bar f + \r c^4,
}}
from which an identification with couplings $u$ and $a$ can be obtained:
\eqn\X{\eqalign{
	\r = \frac{1}{16}\frac{u^2}{a} \qquad   c^2 = -4\frac{a}{u}.
}}
The corresponding $\b$-functions are then given by
\eqn\EQNBETAC{\eqalign{
	\b_{c^2} &= 
		4 - 8 c^2\l^2
}}
\eqn\EQNBETARHO{\eqalign{
	\b_\r &= 
		-\frac{\r}{c^2}\left( c^2- 16c^2{\l}^2 +4 \right).
}}

A close look at \EQNBETAC\ shows that the presence of curvature, parametrized by $\l$, has a stabilizing effect on the radius $c^2$. 
The corresponding $\b$-function vanishes for
\eqn\EQNVANISH{\eqalign{
	c^2 = \frac{1}{2} \l^{-2} =  \frac{1}{2} \kappa.
}} 
Indeed, a spherical 2-brane would be characterized by a finite radius $c$ proportional to $\sqrt\kappa=\l^{-1}$, which is expected from the WZW model. Thus the deformation of the flat background prevents the spherical 2-brane from collapsing. But the condition for vanishing $\b_{c^2}$ still depends on $\l$, which itself is driven by its RG flow and increases. 

\vskip 5mm
\centerline{
\epsfxsize=40ex\epsfbox{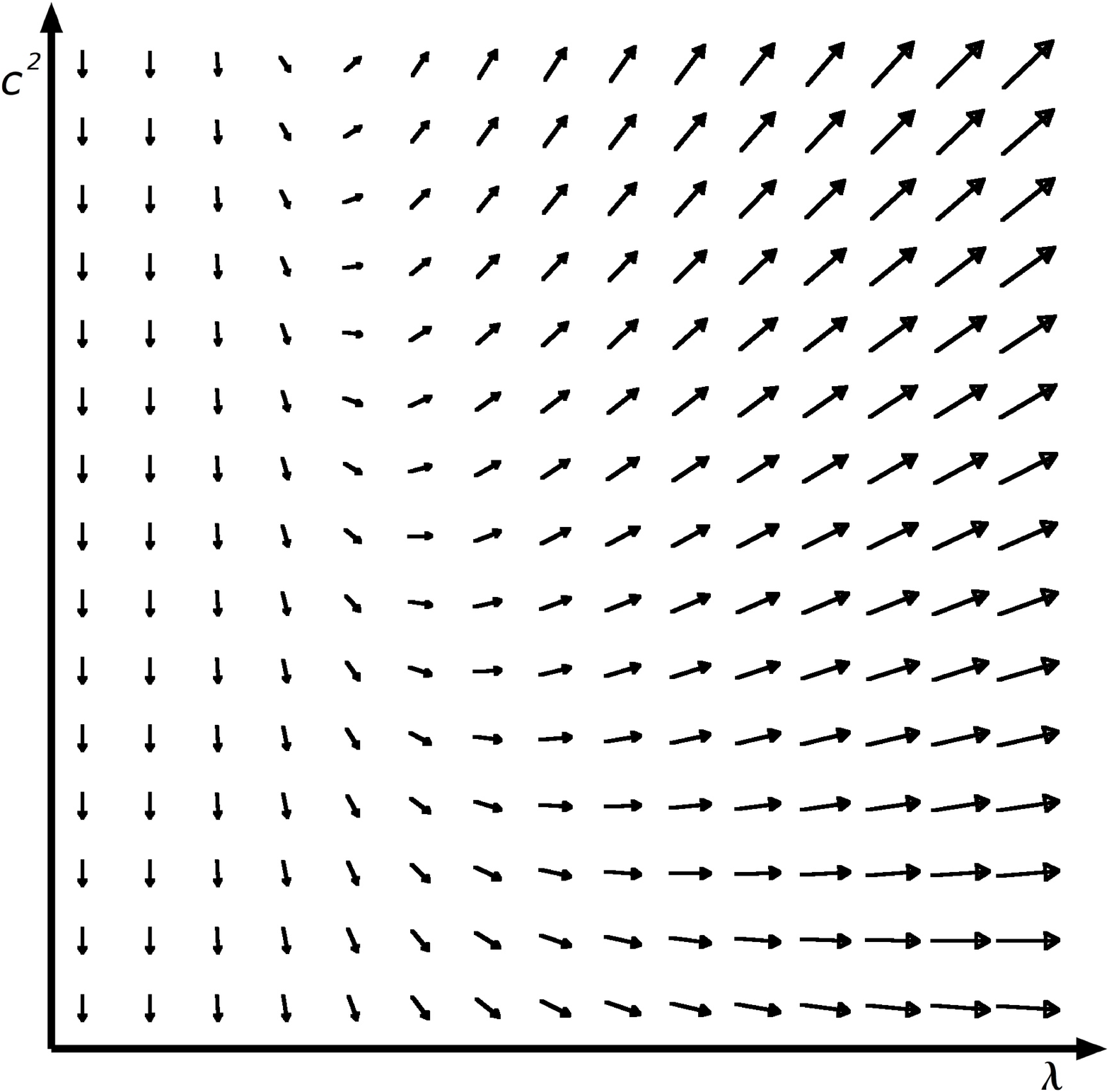}}
\caption{1.7cm}{Fig. 1: The projection of the flow diagram to the $(\lambda,c^{2})$-plane in the parameter region where condensation to a D2-brane starts. Only the presence of non-vanishing $\l$ enables a flow towards finite $c$. The flow can only be trusted for small $\l$.}
\vskip 5mm

For small $\lambda$, where this approximation is valid, $\beta_\rho$ is negative and stays negative after substituting \EQNVANISH. Therefore the coupling $\r$ will increase and trigger a tachyon condensation process.

The perturbative $\beta$-functions on the 3-brane suggest that $c^2 = \frac{1}{2} \l^{-2}$ is not the endpoint of the flow. However, while $\l$ evolves along its RG trajectory, $\r$ increases at a much higher rate. Large $\rho$ on the other hand suggests that the perturbative treatment of the flow on the 3-brane is not applicable any more. Rather than trying to follow the flow all along the RG trajectory it is more reasonable to investigate the conjectured endpoint, a spherical 2-brane.

Note that the original couplings $a$ and $u$ both become infinite, so that the condensation process looks quite like usual tachyon condensation with a configuration with vanishing Zamolodchikov metric as endpoint. But the transformation into $(\r,c^2)$-coordinates reveals, that this condensation is not quite as simple, due to the presence of the non-local coupling $\l$. The submanifold given by this equation is lower-dimensional but curved, a phenomenon which is impossible to observe when tachyon condensation in the presence of only local couplings is studied.

The major drawback of the derivation peresented here is the treatment within perturbation theory while only retaining a finite set of couplings. Exact results are out of reach with this method, and including higher perturbative corrections in the calculations significantly increases the necessary efforts. In order to substantiate the above results we will present another check for the conjectured end point of the renormalization group flow and the existence of a spherical 2-brane with the same qualitative properties in the next section.

\newsec{Stability of the two-brane}

Motivated by the results of the previous section we want to check, if a two-brane is perturbatively stable. We start with a boundary path-integral localized on $X_b^2=c^2$. This constraint is compatible with the group symmetries of $SU(2)$ and describes spherical 2-branes, which are known to be stable. In order to insert this constraint into the action we expand it in the fields $f$ and $\bar f$ to lowest order in $\l$. Moreover we assume that $f^3$ is of the order of the radius $c$ of the 2-sphere, and the other coordinates are small compared to $c$. Explicitly,
\eqn\EQNFIII{\eqalign{
	f^3_n &= - \frac{1}{2c} \left[f^\a f_\a + 2 f^\a\bar f_\a\right]_n,  
}}
where the rhs is projected on the (positive) momentum $n$, and the index $\a$ runs over the directions $\left\{1,2\right\}$. 

Substituting \EQNFIII\ into the action generates several new vertices. In particular the 3-vertex is removed and the 4-vertex shows a much more complicated structure. The interaction consists of several terms, proportional to different combinations of $\l$ and $c$. It is of the form
\eqn\EQNIIBRANE{\eqalign{
	\oint\beta(X_b) &= \frac{\l}{c}A + \frac{s}{c^2}B + \l^2 C.
}} 
The vertices $A,B,C$ are written down explicitly in the appendix.

We want to check, if this action leads to a conformal fixed point describing a  2-brane. Due to the complexity of the generated non-local interaction terms, it is hard to show conformal invariance directly. Therefore we restrict ourselves to the investigation of tachyonic instabilities. For this we need to show that no tachyon is present, neither due to the measure nor due to counter-terms arising at the quantum level. A tachyon would destabilize the 2-brane and initiate a further condensation.

Such information is contained in the various counter-terms appearing in the renormalization procedure. Furthermore the various $\b$-functions should vanish for the theory to be scale-invariant. For this one needs to know the logarithmically divergent counter-terms. More-than-logarithmic divergences tell us, if certain couplings can be set identically to zero in a consistent way. 

For example, we might set a certain coupling $g$ to zero (in an adequate theory). Renormalization then might make it necessary to add a counter-term which excites the coupling $g$. Still, it could be possible to set the renormalized coupling $g_{\rm ren}$ to zero as a renormalization condition. This is then an arbitrary choice and cannot have much physical meaning; it should rather be viewed as a kind of fine tuning of the theory. However, if all counter-terms  vanish, $g=0$ is a solution of the string field theory action.

This is exactly the situation we encounter in our theory \EQNIIBRANE. Due to the complexity of the 4-vertex, the calculation could be done only for vanishing tachyon. However, this is enough to see if tachyonic modes destabilize the 2-brane. According to general scaling arguments, the $\b$-function for $u$ is always proportional to $u$.\foot{As the calculations are done in the limit $u\to 0$ it is impossible to obtain an expression for $\b_u$. This limit involves some care in the regularization of the theory. In particular, the appearance of the correct combinations of $u$ and $R$ (the radius in the disk, which has been set to 1) must be restored in order see the behavior of $\b_u$. The scaling then forces the logarithmic divergences to be proportional to $u$.} 
Hence setting $u=0$ makes its $\b$-function vanish. In order to decide if this condition is 
just fine tuning or has physical relevance,
we need to know the counter-terms. For the 2-point function at vanishing external momentum, they are\foot{The counter-terms have been calculated in the same way as in the previous section. Again we find, that the free field normal ordering prescription can be consistently implemented and is therefore justified a posteriori.}
\eqn\EQNIIDDIV{\eqalign{
	\Sigma^{(2)}(p=0,\Lambda)	
		= &\Bigl [\Lambda\ln\Lambda+(\g-1)\Lambda\Bigr ]
			\left\{\frac{4\l^2}{c^2s^3} -\frac{8}{3}\frac{\l^2}{s^2c^2} - \frac{38}{sc^4}\right\}\cr
		&+\ln\Lambda\left\{{\rm terms\;proportional\; to\;}u\right\} \cr
}}

Most remarkably the more than logarithmically divergent counter-terms are not independent of each other. They arrange themselves in a way so that they all appear with the same factor. Therefore it is possible to remove them altogether by imposing one single condition, adjusting the value of the radius $c$. Setting for example $s=1$ gives
\eqn\EQNLOC{\eqalign{
	c^2 &= \frac{57}{2}\l^{-2} = \frac{57}{2}\kappa .
}}
Of course, the numerical factor is still modified by wave function renormalization, which has not been taken into account here.

The logarithmic part of \EQNIIDDIV\ contains the 2-loop contribution for $\beta_u$. To prove conformal invariance at the 2-loop level one ought to establish the absence of counter-terms for the other couplings as well, which we have not obtained here. Rather we want to stress the absence of higher-than-logarithmic divergences in the counter-terms after imposition of localization to \EQNLOC\ as a check for the claim on the end point of RG-flow of the decaying $3$-brane.

It is tempting to view the RG-behavior of our model as realization of 't Hooft's naturalness principle, albeit in a different context than confining gauge theories, for which it was originally formulated \tHooftXB. Natural theories do not need fine-tuning of the couplings in order to cancel counter-terms; therefore, small parameters stay small under a change of scale, which is a property shared by our model. The physical picture behind is, that small couplings are preferred, when their vanishing increases a symmetry. One could speculate about symmetry enhancement in the above model. Reversing the argument would imply that some symmetry exists which fixes $c^2$ to a certain value. This is reminiscent of the quantization of radii of D-branes in $SU(2)$ and nourishes hope that higher order perturbation theory could reveal a D-brane potential capable of describing localization on quantized D-branes.

\newsec{Discussion and interpretation}

Although we have applied the open-closed string correspondence developed in \BaumgartlIY\ to a specific example the qualitative features observed here should be rather generic since apart from the numerical values not much depended on the details of the group manifold in question. Given the highly symmetric set-up one might hope that some of the phenomena discussed here within perturbation theory could be established exactly at least for some simple processes. In particular within the perturbative approximation utilized here we are not able to see all D2-branes corresponding to conjugacy classes of the group. Rather we only see the `biggest' 2-brane. This should be related to the fact that we worked in the large radius regime. Pushing the perturbation in $\l$ further it is conceivable that additional fixed points appear which describe `smaller' conjugacy classes, but in order to see these much more powerful methods are needed. More interestingly it would be worthwhile investigating, if non-symmetry preserving branes exist in these models. 

Although we have observed the absence of divergences in the 2-brane theory by brute force computation, there may well be symmetry arguments that imply finiteness of the loop correction. It would be interesting to know if such a symmetry exists, in particular in view of a non-perturbative approach to these models. 

Finally the presence of the non-local coupling $\lambda$ deserves an interpretation in terms of open string field theory. We expect that it should be related to some condensate of the open string degrees of freedom. A better understanding should be useful in view of the open-closed string correspondence.

\newsec{Acknowledgments}

The authors would like to thank 
M. Gaberdiel, 
M. Kiermayer,
D. Laenge,
B. Jurco,
J. Pawlowski,
G. Policastro,
A. Recknagel,
M. Schnabl,
S. Shatashvili
and
A. Tseytlin
for helpful discussions. This work was supported by SFB-375 and SPP-1096 of the DFG.

\newpage

\newsec{Appendix: the three-dimensional theory}

Here we collect all the formulas which allow to derive the explicit expressions in \SECBOUNDARYACTION.

\subsec{The action}

Let $T_\m$ be the generators of $SU(2)$.
We define the operators
\eqn\X{\eqalign{
	{\rm ad}_f = \left [f^\m T_\m, \cdot\;\right ] \qquad {\rm Ad}_h = h\;\cdot\;h^{-1}
}}
and derive (with $\w(\theta) \equiv 2\sqrt{f^\mu(\theta)f_\mu(\theta)}\,\,\,$)
\eqn\X{\eqalign{
	{\rm Ad}_h  &= {\rm id} + \frac{i\sin \lambda\w}{\w}{\rm ad}_f  
			+ \frac{\cos\lambda\w-1}{\w^2}{\rm ad}_f^2 \cr
	\d hh^{-1} &= \left [i\lambda {\rm id} + \frac{\cos \lambda\w-1}{\w^2}{\rm ad}_f 
		+ i\frac{\sin \lambda\w-\lambda\w}{\w^3}{\rm ad}_f^2\right ]\delta f^\m T_\m\cr
	h^{-1}\d h &= \left [i\lambda {\rm id} - \frac{\cos \lambda\w-1}{\w^2}{\rm ad}_f 
		+ i\frac{\sin \lambda\w-\lambda\w}{\w^3}{\rm ad}_f^2\right ]\delta f^\m T_\m.\cr
}}
With these preparations the action can be obtained exactly in these coordinates. An expansion in the perturbation parameter $\l$ up to order $\l^3$ yields the expressions \EQNAC\ and \EQNACII\foot{A normalization of boundary integrals has been used, which absorbs factors of $2\pi$ in a convenient way.}.

\subsec{The Jacobian}

Here we present details about the calculation of the Jacobian as advocated in \SECBOUNDARYACTION. Unlike for the action, it is not possible to obtain an explicit expression for arbitrary $\l$, but a perturbative expansion is possible. 
Let us first focus on the determinant of the matrix $J_{11} = \frac{[\d hh^{-1}]^+}{\d f}$. In components it can be expressed as
\eqn\X{\eqalign{
	(J_{11})_{nm}^{\m\n} &= 
		i\l \d_{nm}^{\m\n} - 2i\oint d\t e^{i(n-m)\t}
			\left [{\epsilon_\rho}^{\m\n} f^\rho(\t)\frac{{\rm d}}{{\rm}d\lambda} + 2 \epsilon^{\m \l \r}{\epsilon_\r}^{\k\n}f^\l (\t)f^\k(\t)\right ]\cal{A}(\lambda)\cr 
			\cal{A}(\lambda)&=
			\frac{\sin\lambda\omega-\lambda\omega}{\omega^3}
}}
The functions $f(\t)$ are given by the holomorphic function $f(z)$ with coordinates restricted to the boundary $z=e^{i\t}$. As $f$ has no zero mode, the integral can only be non-zero when $m>n$. In particular only the very first term contributes to the trace of $J_{11}$. Higher powers of $J_{11}$ contain terms $\oint e^{i(n-p_1)\t_1}\oint e^{i(p_1-p_2)\t_2}\cdots\oint e^{i(p_k-m)\t_{k+1}}$ with $n<p_1<\cdots<p_k<m$. Under the trace these terms vanish again. Therefore (we suppress irrelevant factors coming from tracing over space indices)
\eqn\X{\eqalign{
	\Tr\,J_{11}^n &= (i\l)^n \Tr\, 1.
}}
Using the expansion of the determinant in traces, 
\eqn\X{\eqalign{
	\ln\Det\frac{J_{11}}{i\l} = \Tr \int \frac{ds}{s}e^{-s\frac{J_{11}}{i\l}},
}}
we get
\eqn\X{\eqalign{
	\Det \frac{J_{11}}{i\l} = 1.
}}
For the computation of $J_{22}$ we expand
\eqn\X{\eqalign{
	h\d\bar h\bar h^{-1}h^{-1}
	&= \sum_{i=0}^8 \d b_i\cr
	\d b_0 = i\l\d\bar f &\qquad 
		\d b_1 = \bar Z_1{\rm ad}_{\bar f}\d\bar f\cr
	\d b_2 = \bar Z_2{\rm ad}_{\bar f}^2\d\bar f &\qquad 
		\d b_3 = i\l Z_3{\rm ad}_f\d\bar f\cr
	\d b_4 = Z_3\bar Z_1{\rm ad}_f{\rm ad}_{\bar f}\d\bar f &\qquad 
		\d b_5 = Z_3\bar Z_2{\rm ad}_f{\rm ad}_{\bar f}^2\d\bar f\cr
	\d b_6 = i\l Z_1{\rm ad}_f^2\d\bar f &\qquad
		\d b_7 = Z_1\bar Z_1{\rm ad}_f^2{\rm ad}_{\bar f}\d\bar f\cr
	\d b_8 = Z_1\bar Z_2{\rm ad}_f^2{\rm ad}_{\bar f}^2\d\bar f.&
}}
with the abbreviations
\eqn\X{\eqalign{
	Z_1 &= -\frac{\l^2}{2} + \frac{\l^4 \w^2}{24} - \frac{\l^6\w^4}{720} + {\cal O}(\l^7) \cr
	Z_2 &= -\frac{i\l^3}{6} + \frac{i\l^5\w^2}{120} + {\cal O}(\l^7) \cr
	Z_3 &= i\l + \w^2 Z_2.
}}
The functional matrices, which enter the determinant are
\eqn\X{\eqalign{
	\left [\frac{\d b_i}{\d \bar f}\right ]^-_{(\m n)(\n m)},
}}
where the upper index `$-$' indicates projection on the antiholomorphic modes.
Due to $m\le n$, they are all upper triangular matrices, hence the determinant is just the product of the diagonal entries. We re-write it in the following way:
\eqn\EQNDET{\eqalign{
	\Det \frac{J_{22}}{i\l} \equiv e^{\Tr \ln  B_{nm}^{\m\n}},
}}
where
\eqn\X{\eqalign{
	B_{nm}^{\m\n} = \d^{\m\n}_{mn} &- \frac{4i}{\l}\oint Z_3\bar Z_1f^\l\bar f^\k \epsilon^{\k\n\r}\epsilon^{\r\l\m}\cr
					&-\frac{8}{\l}\oint \left [Z_3\bar Z_2 f^\l \bar f^\k\bar f^\r
						+ Z_1\bar Z_1 f^\l f^\k\bar f^\r\right ]
							\epsilon^{\r\n\s}\epsilon^{\k\s\tau}\epsilon^{\l\tau\m}\cr
		&+\frac{16i}{\l}\oint Z_1\bar Z_2f^\l f^\k\bar f^\r\bar f^\s 
		\epsilon^{\s\n\tau}\epsilon^{\tau\r\w}\epsilon^{\w\k\xi}\epsilon^{\xi\l\m}.
}}
Next we expand the logarithm \EQNDET\ and derive an expression for the contribution to the action. A straight forward calculation reveals
\eqn\X{\eqalign{
	I_{\rm Jacobian} &= 4\l^2 \sum_{n>0}f_n^\m f_{n\m} + {\cal O}(\l^4). 
}}
The lowest order of the Jacobian, which modifies the action, has therefore the form of a tachyon interaction.

\newsec{Appendix: the two-dimensional theory}

In this section $f$ denotes the fields of the 2-dimensional theory, $f^\a=(f^1, f^2)$.
Using the constraint \EQNFIII\ together with the three-dimensional action yields a two-dimensional action. Its interaction terms are given by the following expression, supplemented by the standard kinetic term:
\eqn\X{\eqalign{
	I &= \frac{\l}{c}A + \frac{s}{c^2}B + \l^2 C.
}}
The three types of interactions, distinguished through their dependence of combinations of $c$ and $\l$, are given by
\eqn\X{\eqalign{
	A &= \sum_i A_i - \sum_i \bar A_i\cr
	B &= B_1+B_2+B_3+\bar B_3\cr
	C &= C_1+C_2+\bar C_2
}} 
with
\eqn\X{\eqalign{
	% A was V,
	% B was K,
	% C was R
	A_1 &= \frac{1}{2}\sum_{a=1}\sum_{d=2}\sum_{e=1}^{d-1}(d-a) \epsilon_{\a\b}f^\a_a\bar f^\b_{a+d}{f_e}_\g f_{d-e}^\g 
		\cr
	\bar A_1 &= A_1^*\cr
	A_2 &= \sum_{a=1}\sum_{c=1}\sum_{d=2}(d-a)\epsilon_{\a\b}f^\a_a\bar f^\b_{a+d}{f_{c+d}}_\g\bar f^\g_c
		\cr
	\bar A_2 &= A_2^* \cr
	A_3 &= \frac{1}{4}\sum_{a=1}\sum_{b=1}\sum_{g=1}^{a+b-1}(a-b)\epsilon_{\a\b}f_a^\a f_b^\b \bar f_{g\g}\bar f^\g_{a+b-g}
		\cr
	\bar A_3 &= A_3^* \cr
	A_4 &= \frac{1}{2}\sum_{a=1}\sum_{b=1}\sum_{c=1}(a-b)\epsilon_{\a\b} f^\a_a f^\b_b f_{c\g}\bar f_{a+b+c}^\g
		\cr
	\bar A_4 &= A_4^* \cr
	B_1 &= \frac{1}{4}\sum_{m=2}\sum_{a=1}^{m-1}\sum_{b=1}^{m-1} m f_{a\a}f^\a_{m-a}\bar f_{b\b}\bar f^\b_{m-b}
		\cr
	B_2 &= \sum_{m=2}\sum_{a=1}^{m-1}\sum_{b=1}^{m-1} m \bar f_{a\a} f^\a_{m-a} f_{b\b}\bar f^\b_{m-b}
		\cr
	B_3 &= \frac{1}{2}\sum_{m=2}\sum_{a=1}^{m-1}\sum_{b=1}^{m-1} m f_{a\a}f^\a_{m-a} f_{b\b}\bar f^\b_{m-b}
		\cr
	C_1 &= -\sum_{a,b,c,d=1}\frac{(a-b)(c-d)}{c+d} f^\a_a\bar f_{c\a} f^\b_b\bar f_{d\b} \d_{a+b,c+d} \cr
	C_2 &= \frac{1}{3}\sum_{a,b,c,d=1}(c-a-b)\bar f^\a_a\bar f_{b\a}\bar f_{c\b} f^\b_{a+b+c}\cr
	\bar C_2 &= C_2^*.
}}
A contribution from the 3d measure has been not included, since after imposing the constraint it would appear as a cosmological constant. For the $\b$-functions this is taken into account anyway as part of the tachyon couplings.

%%%%%%%%%%%%%%%%%%%%%%%%%%%%%%%%%%%%%%%%%%%%%%%%%%%%%%%%
\listrefs
\bye